\documentclass[twocolumn,pre]{revtex4}

\usepackage{graphicx}
\usepackage{dcolumn}
\usepackage{bm}
\usepackage{amsmath}
\usepackage{graphics}
\begin{document}

\title{A model of HIV budding and self-assembly, role of cell membrane}

\author{Rui Zhang and Toan T. Nguyen}
\affiliation{School of Physics, Georgia Institute of Technology,
837 State Street, Atlanta, Georgia 30332-0430}

\date{\today}

\begin{abstract}

Budding from the plasma membrane of the host cell is an
indispensable step in the life cycle of the Human Immunodeficiency
Virus (HIV), which belongs to a large family of enveloped RNA
viruses, retroviruses. Unlike regular enveloped viruses,
retrovirus budding happens {\em concurrently} with the
self-assembly of retrovirus protein subunits (Gags) into spherical
virus capsids on the cell membrane. Led by this unique budding and
assembly mechanism, we study the free energy profile of retrovirus
budding, taking into account of the Gag-Gag attraction energy and
the membrane elastic energy. We find that if the Gag-Gag
attraction is strong, budding always proceeds to completion.
During early stage of budding, the zenith angle of partial budded
capsids, $\alpha$, increases with time as $\alpha \propto
t^{1/3}$. However, when Gag-Gag attraction is weak, a metastable
state of partial budding appears. The zenith angle of these
partially spherical capsids is given by
$\alpha_0\simeq(\tau^2/\kappa\sigma)^{1/4}$ in a linear
approximation, where $\kappa$ and $\sigma$ are the bending modulus
and the surface tension of the membrane, and $\tau$ is a line
tension of the capsid proportional to the strength of Gag-Gag
attraction. Numerically, we find $\alpha_0<0.3\pi$ without any
approximations. Using experimental parameters, we show that HIV
budding and assembly always proceed to completion in normal
biological conditions. On the other hand, by changing Gag-Gag
interaction strength or membrane rigidity, it is relatively easy
to tune it back and forth between complete budding and partial
budding. Our model agrees reasonably well with experiments
observing partial budding of retroviruses including HIV.

\end{abstract}

\maketitle

\newcommand{\bq}{\begin{eqnarray}}
\newcommand{\eq}{\end{eqnarray}}
\newcommand{\arcsinh}{\mathop{\mathrm{arcsinh}}}

\section{Introduction}

The Human Immunodeficiency Virus (HIV) is famous for its ability
to induce Acquired Immunodeficiency Syndrome (AIDS). It belongs to
a large family of enveloped RNA viruses, retroviruses.
Retroviruses are characterized by the unique infection strategy of
reverse transcription, in which the genetic information flows from
RNA back to DNA (therefore the name ``retro")~\cite{Coffin}.
Budding is an indispensable step in the retroviral life
cycle~\cite{Sundquist04,Freed04}. After the major retroviral
structural protein, Gags, are synthesized inside the host cell,
they are transported to the cell membrane and self-assemble into
spherical protein shells called ``capsids", with viral RNA genome
and other auxiliary viral proteins packaged inside. At the same
time, these capsids, enveloped by the cellular membrane, have to
bud out of the membrane to target other host cells.  In other
words, budding and assembly of retroviruses happen {\em
concurrently} on the cell membrane.
\begin{figure}[h]
\begin{center}
\includegraphics[width=0.4\textwidth]{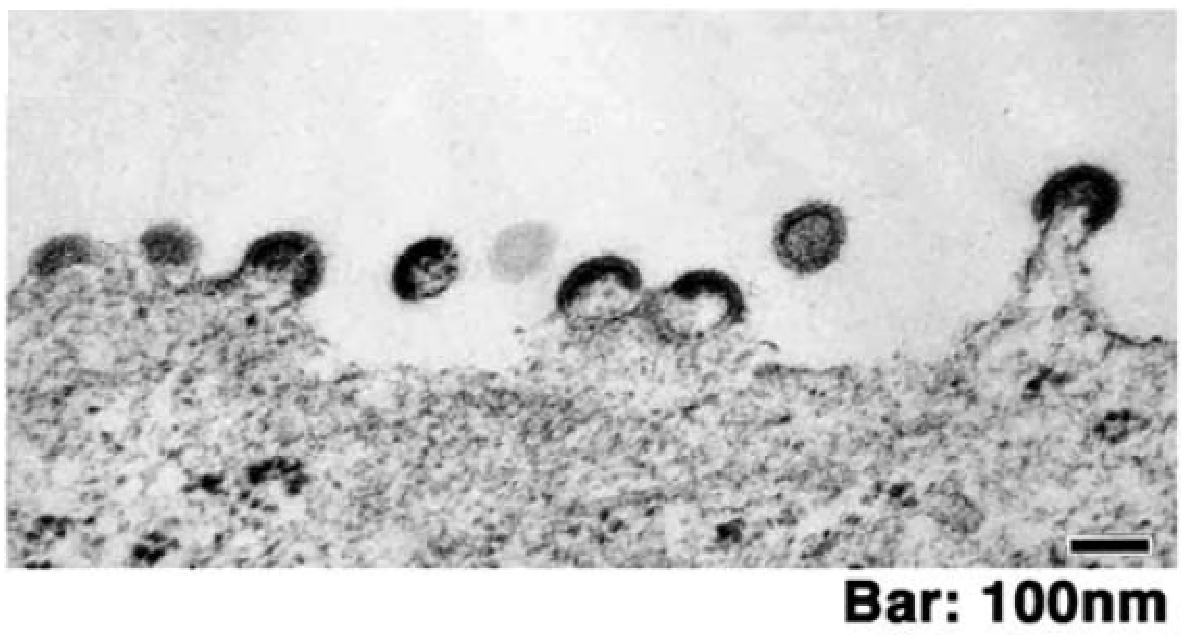}
\end{center}
\caption{ Electron microscopic image of partial budding of HIV-1
viruses (reprinted from~\cite{Sundquist04}). } \label{fig:budexp}
\end{figure}
Despite a large body of experiments done within the last decade,
the biological pathway and mechanism of retroviral budding have
still not been fully understood~\cite{Sundquist04,Freed04}. One
important unexplained observation is that viral budding can be
inhibited partially or completely by changing the cell environment
or mutating the late domains of the Gag proteins. In these
situations, capsids are only partially formed and stuck on the
membrane (Fig.~\ref{fig:budexp}). Motivated directly by this
partial budding phenomenon, in this paper, we propose a physical
model to study HIV (and retroviruses in general) budding and
assembly on the elastic membrane. Physically, this situation is
interesting because it provides a unique two dimensional
self-assembly mechanism in which the membrane elastic energy plays
an important role, since assembly is always accompanied with
budding. Biologically, understanding the physical mechanism of HIV
budding and assembly is certainly important toward understanding
the HIV life cycle. It is also important in the
light of recent effort from the virology community to develop
assembly-oriented anti-viral therapy.

\begin{figure}[h]
\begin{center}
\includegraphics[width=0.4\textwidth]{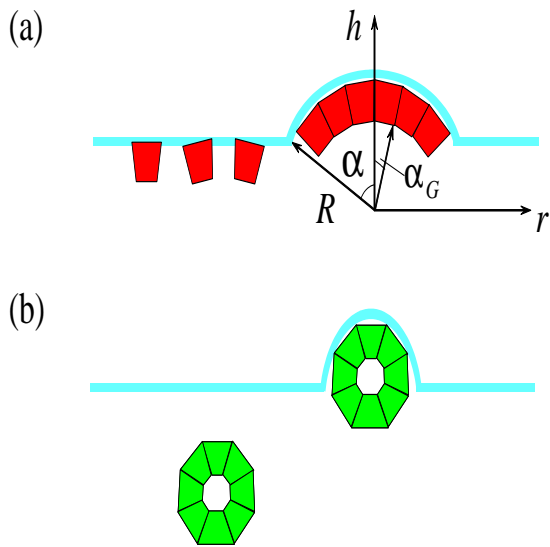}
\end{center}
\caption{(Color online) Schematic illustrations of two different
types of virus budding. (a): budding of retroviruses: capsid
proteins (Gags) are first attracted to the membrane, then
self-assemble and bud on the membrane at the same time. This is
the system studied in this paper. (b): budding of regular
enveloped viruses: capsid proteins first self-assemble into
complete capsids inside the cell, then bud on the
membrane~\cite{Gelbart}. (a) also shows the cylindrical coordinate
system $(r, h, \phi)$ used in the model (the polar angle $\phi$ is
not shown).} \label{fig:gags}
\end{figure}

Budding of regular enveloped viruses was studied theoretically by
several authors (TDGB) in Ref.~\cite{Gelbart,Deserno03,Deserno04}.
However, the viral budding pathway and the physical model studied
by TDGB is qualitatively different from retroviral budding we
study in this paper. For regular enveloped viruses, viral capsids
are fully assembled inside the cell
\cite{ZlotnickJMB07, ChandlerBJ06,ShklovskiiKinetic07}.
After that, they are transported to the cell membrane,
bind to the viral spike proteins
(embedded in the cell membrane) and then bud out through the cell
membrane (see Fig.~\ref{fig:gags}b). Therefore, capsids formation
and budding out of the membrane are two separate processes. And
the main driving force of budding is the capsid-membrane
attraction (mediated by embedded spike proteins).

Budding of retroviruses follows a completely different pathway.
Various TEM and X-ray tomography experiments suggest that
retroviral capsids are assembled from Gag proteins on the cell
membrane and bud out of the cell {\em concurrently}
~\cite{Sundquist04,Freed04}. Based on these experiments, we study
a different model for HIV (and retrovirus in general) budding and
assembly shown in Fig.~\ref{fig:gags}a. In this new model, we
assume retroviral capsids are assembled from membrane-bound Gags
only, neglecting the possibility that Gags from the interior of
the cell may participate. In other words, the Gag-membrane
attraction is strong such that Gags always bind to the membrane.
This assumption is supported by various experimental observation
where budding is completely inhibited (no capsids are formed) but
Gags are found in abundance at the cell membrane~\cite{Lingappa07}
. In contrast to the TDGB model, the primary driving force of our
retroviral budding is the short range attraction between these
membrane-bound Gag proteins. This correlates well with
experimental fact that point mutations changing Gag-Gag
interactions affect the degree of viral budding. On the other
hands, spike proteins or virus RNA seem not important for
retroviral budding. In vitro, Gag proteins are directly attracted
to the membrane and they alone are usually sufficient for the
assembly and release of virus-like
particles~\cite{Freed04,Garoff98,Krausslich07}. We therefore
neglect the contribution of all other proteins or RNA  components
of retroviruses in our model.

In this paper, for a given set of parameters (the membrane Gag
concentration, the Gag-Gag interaction, and the cell membrane
bending and stretching rigidity), we study the free energy profile
of budded viral capsids. Two energies are considered explicitly:
first, the elastic energy of the membrane including the bending
and stretching energy; second, the Gag-Gag attraction energy when
a Gag makes contact to the other Gag (see Fig.~\ref{fig:gags}a).
Since the elastic energy scale is much larger than
k$\mathrm{_B}$T, for example, the bending rigidity of normal
membranes is about $20$ k$\mathrm{_B}$T, thermal fluctuations are
higher order corrections and neglected in the theoretical
treatment. Focusing on the budding process, we also assume that
the Gag-Gag interaction is strong enough such that the entropic
cost of bringing free Gags to the capsid can be ignored. For
simplicity, we assume the shape of the capsid together with the
membrane attached to it is (partially) spherical with radius $R$
(Fig.~\ref{fig:gags}a). The size of a capsid is then characterized
by the zenith angle $\alpha$ at its edge, the smallest being the
angle of a single Gag protein, $\alpha_G$ (Fig.~\ref{fig:gags}a).
Since $\alpha_G$ is very small ($\alpha_G=0.03$ for a typical HIV
capsid containing 5000 Gags), we take $\alpha_G\rightarrow 0$ in
the theoretical consideration and treat $\alpha$ as a continuous
variable. As budding proceeds to completion, $\alpha$ increases
from $\alpha_G$ to $\pi$. When $\alpha=\pi$, the capsid actually
leaves the membrane through membrane fission. In this paper, we do
not consider this fission process and thus, in our terminology,
complete budding always means $\alpha\rightarrow\pi$. To simplify
the calculation, we employ an scaling description where we neglect
the variation in the degree of viral budding and assume all
capsids have the same average zenith angle $\alpha$.

\begin{figure}[h]
\begin{center}
\includegraphics[width=0.52\textwidth]{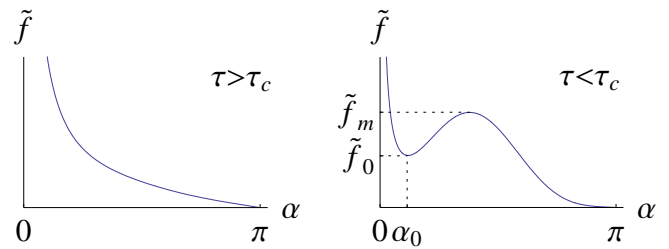}
\end{center}
\caption{The schematic illustration of the total free energy
density as a function of the capsid size $\alpha$. The left and
right profiles correspond to strong or weak Gag-Gag attraction
respectively. Here $\tau$ is the line tension of the rim of a partially budded
capsid. $\tau$ is proportional to the strength of Gag-Gag attraction. $\tau_c$ is
the threshold line tension at which the local minimum at $\alpha_0$
appears.} \label{fig:free}
\end{figure}

Our main result is shown in Fig.~\ref{fig:free}. The key parameter
is the strength of Gag-Gag attraction which can be adjusted
experimentally by mutating of Gags, complexing of Gags with
other molecules or by changing pH or salinity
of the cell cytoplasm near the membrane
\cite{Sundquist04,ReinPNAS01}. In a partially budded
capsid, the line tension $\tau$
of the rim of a capsid is directly proportional to this interaction
Gag-gag interaction.
When the Gag-Gag attraction is strong (or when $\tau$ is greater
than a threshold value $\tau_c$), as in the normal biological
conditions of HIV, budding always proceeds to completion, i.e.,
$\alpha\rightarrow\pi$ (the left panel of Fig.~\ref{fig:free}). At
early stage of budding, the size of a partially budded capsid
increases very slowly with time:
\bq
\alpha(t) \approx (t/\tau_{diff})^{1/3}
\eq
where the time scale $\tau_{diff}$ depends on the lateral mobility
of Gag, the radius of the capsid and the initial concentration Gag
(see Eq. (\ref{tauDiff})). On the other hand, when the Gag-Gag
attraction is weak ($\tau < \tau_c$), for example, after mutation of the late
domains of the Gag protein, partial budding appears as a
metastable state at the capsid size $\alpha_0$ (the right panel of
Fig.~\ref{fig:free}). In this case, the free energy barrier can be
much larger than $k_BT$ and budding is kinetically trapped at
$\alpha_0$. Using a linear approximation, we find \bq
\alpha_0\simeq\sqrt[\uproot{10}4]{\frac{\tau^2}{\kappa\sigma}},\eq
where $\sigma$ and $\kappa$ are the surface tension and bending
rigidity of the membrane. 

The energetics of HIV budding and assembly is studied both
analytically and numerically in this paper.
Analytically, the complete scaling behaviors of the free energy
density profile in asymptotic limits of ``soft" and "stiff"
membranes are calculated (the meaning of "soft" and "stiff" membrane
will be clear in later sections). In all cases, they agree with
the numerical result well. However, the numerical result gives a
complete solution to the problem including nonlinear regimes where
the analytical result is normally not available. The inequility
$\alpha_0<0.3\pi$ is found to always hold from the numerical calculation
without any approximations.
%
%

It is worth to point out that
budding in our model can be considered as a consequence of the
inhomogeneity of the membrane if one considers Gags as a part of
the membrane. In this sense, our work is related to J\"{u}licher
and Lipowsky's works on domain-induced budding of
vesicles~\cite{Lipowsky93,Lipowsky96}. However, in their papers,
the inhomogeneity was introduced through two kinds of lipids which
do not carry a given curvature like our Gags. Domain-induced
budding is a consequence of demixing of these different molecules.
As a result, their budding happens in a much larger length scale
(comparable to the size of the vesicle) where only two phases
coexist, one budded out from the other. While in our case, we
consider budding at a much smaller length scale (a typical HIV-1
virus particle is about 130 nm in diameter, which is a hundred
times smaller than the size of a host cell) and actually have a
multi-phase coexistence since there are more than one capsid on
the membrane.

This paper is organized as follows. In Sec.~\ref{sec:model} , we
introduce the physical model of HIV budding and assembly. We then
discuss the analytical solution to the elastic energy of the
membrane in Sec.~\ref{sec:elastic} and to the total free energy
density in Sec.~\ref{sec:analytical}. The numerical result is then
provided and compared to the analytical results in
Sec.~\ref{sec:numerical}. After we get the complete theoretical
result, we discuss budding kinetics and make connections to
experiments in Sec.~\ref{sec:exp}. We finally conclude in
Sec.~\ref{sec:conclusion}. In this paper, the term ``capsid" is
used for both partial and complete spherical shells of viral
proteins. The meaning should be clear from the context.

\section{The elastic model of HIV capsid budding and self-assembly\label{sec:model}}

Let us consider a membrane-capsids system in which the
concentration of Gags on the membrane, $c_G$, is fixed. We assume
all capsids assembled by Gags have the same average zenith angle
$\alpha$ (see Fig.~\ref{fig:gags}a), and an average concentration,
$n$. $n$ is related with $\alpha$ by the conservation of mass of
Gags:
\bq n = c_G \frac{A(\alpha_G)}{A(\alpha)} =
c_G\frac{1-\cos\alpha_G}{1-\cos\alpha}, \label{eq:cG} \eq where
\bq A(\alpha)=2\pi R^2 \int_0^{\alpha}\sin\theta d\theta=2\pi
R^2(1-\cos\alpha)\label{eq:A}\eq
is the area of a capsid with zenith angle $\alpha$, and $\alpha_G$
is the zenith angle of a single Gag (see Fig.~\ref{fig:gags}a).
Within this average description, it is convenient to think that
the whole membrane surface is divided into identical cells, each
contains a single capsid. The average size of these approximately
circular cells, $d$, is given by the condition
\bq \pi (d/2)^2n=1. \label{eq:d}\eq

Generically, the free energy density of the membrane-capsid system
can be written as
\bq f=n\varepsilon=n(\varepsilon_m+\varepsilon_c), \label{eq:f}
\eq
where $\varepsilon$ is the free energy of one membrane cell. It
includes two parts: the elastic energy of the membrane,
$\varepsilon_m$, and the capsid energy $\varepsilon_c$ coming from
the Gag-Gag interaction and the Gag-membrane interaction.

To calculate the elastic energy of the cell membrane, we use the
standard Helfrich model~\cite{Canham,Helfrich} where
$\varepsilon_m$ is the sum of two contributions from the bending
energy and the stretching energy:
\bq \varepsilon_m=\int dS\left[\frac{\kappa}{2}(2H-C_0)^2+\kappa_G
K\right]+\int dS \sigma. \label{eq:Helfrich} \eq
Here the integration with the area element $dS$ is taken over the
membrane surface. $\kappa$ and
$\kappa_G$ are the bending rigidity and Gaussian bending rigidity,
$H$ and $K$ are the mean and Gaussian curvatures, and $C_0$ is the
spontaneous curvature of the membrane surface. Using the
Gauss-Bonnet theorem, one can show that the total Gaussian
curvature of the membrane surface is proportional to the total
area of capsids, in the generic case when $\kappa_G$ takes
different values for the membrane attached to the capsid and the
Gag-free membrane. Since the Gag concentration $c_G$ in our system
is fixed, this term gives a constant in $f$ and can be dropped
from further consideration~\cite{G}. For a given Gag concentration
$c_G$, under our spherical capsid assumption, the shape and the
total area of all capsids are fixed. Therefore the total elastic
energy of the membrane attached to capsids is also constant, and
can also be dropped from consideration. As a result, the
$\alpha$-dependent contribution to $\varepsilon_m$ comes from the
integration over the Gag-free membrane surface only. In this
region, we take the spontaneous curvature to be $C_0=0$,
corresponding to normal lipid bilayer membranes.

In consideration of the single capsid energy $\varepsilon_c$,
since $c_G$ is constant, both the total Gag-membrane interaction
energy and the bulk part of the Gag-Gag interaction energy are
constant. The only $\alpha$-dependent contribution to
$\varepsilon_c$ comes from the rim energy of the capsid, due to
the fact that the coordination number of Gags on the rim
is not as many as Gags inside the capsid. Since the perimeter of
the capsid rim with zenith angle $\alpha$ is $2\pi R\sin\alpha$,
we set
\bq \varepsilon_c=\tau 2\pi R\sin\alpha.\label{eq:Ec} \eq
The proportionality coefficient $\tau$ can be considered as the
``line tension" of the capsid. It is directly proportional to the
strength of the Gag-Gag attraction and can be changed
experimentally by mutations of Gags or by changing pH or salinity
of the cell cytoplasm near the membrane.

To proceed further, we take the ``ideal capsids" approximation
when the distance between capsids is large and the
membrane mediated interaction between them is negligible. Such an
effective long-range interaction is possible because the presence
of the first capsid may change the deformation of the membrane
around the second capsid and provides an effective interacting
energy between the two. 
Qualitatively, this interaction is
negligible when the capsid concentration $n$ is small (the
quantitative condition will be given in the next section). 
Under this non-interacting
capsids approximation, $\varepsilon_m$ comes from the membrane
deformation induced by a \emph{single} capsid. 

The calculation procedure to find the free energy profile
$f(\alpha)$ is as follows. We first minimize the membrane elastic
energy $\varepsilon_m$ with respect of all possible membrane
shapes for any given capsid size $\alpha$. Here it is convenient
to use a cylindrical coordinate system $(r, h, \phi)$ as shown in
Fig.~\ref{fig:gags}a (the azimuthal angle $\phi$ is not shown).
With our assumption of (partial) spherical capsids, the membrane
profile is independent on $\phi$. As a result, one can use either
the function
 $h(r)$ or $ r(h)$ to parameterize the membrane.
Correspondingly, the mean curvature and the area element can be
written as~\cite{Kreyszig}
\bq
H(r)&=&\frac{h'(r)^3+h'(r)+rh''(r)}{2r[1+h'(r)^2]^{3/2}},\\
dS&=&r\sqrt{1+h'(r)^2}drd\phi;\\
\mathrm{or} \ \ H(h)&=&\frac{1+r'(h)^2-r(h)r''(h)}{2r(h)[1+r'(h)^2]^{3/2}},\label{eq:H}\\
dS&=&r(h)\sqrt{1+r'(h)^2}dhd\phi,\eq
where $h'(r)=dh/dr$ and $h''(r)=d^2h/dr^2$ are the first and
second derivatives of $h$ with respect to $r$. Similarly,
$r'(h)=dr/dh$ and $r''(h)=d^2r/dh^2$ are the first and second
derivative of $r$ with respect to $h$. Functionally minimizing the
membrane energy $\varepsilon_m$ with respect to membrane shape
$r(h)$ or $h(r)$, one obtains an elastic equation of the membrane
shape, similar to the Euler-Lagrange equation derived from the
least action principle in the classical mechanics. For the shape
parametrization using $r(h)$, $\delta\varepsilon_m/\delta r=0$
leads to the equation:

\begin{widetext}
\bq &&\frac{\kappa}{2 r^2 [1+r'^2]^{9/2}}[-r'^2-3 r'^4-3 r'^6
-r'^8+r r''-3 r r'^4 r''-2 r r'^6 r''+2 r^2 r''^2-11 r^2 r'^2
r''^2-13 r^2 r'^4 r''^2-5 r^3 r''^3\nonumber\\
&&+30 r^3 r'^2 r''^3+4 r^2 r' r^{(3)}+8 r^2 r'^3 r^{(3)}+4 r^2
r'^5 r^{(3)} -20 r^3 r' r'' r^{(3)}-20 r^3 r'^3 r'' r^{(3)}+2
r^3 r^{(4)}+4 r^3 r'^2 r^{(4)}+2 r^3 r'^4
r^{(4)}]\nonumber\\&&+\sigma\frac{1+r'^2-r r''}{
[1+r'^2]^{3/2}}=0,
\label{eq:EL1}
\eq
\end{widetext}
where $r^{(3)}=d^3r/dh^3$ and $r^{(4)}=d^4r/dh^4$ are the third and
forth derivatives of $r$ with respect of $h$. This equation has to be
solved together with the boundary conditions. On the rim of the
partial spherical capsid, the membrane itself and its slope must
be continuous. We have
\bq h(r)|_{R\sin\alpha}=R\cos\alpha,\ \
h'(r)|_{R\sin\alpha}=-\tan\alpha,\label{eq:bchr};\eq or \bq
r(h)|_{R\cos\alpha}=R\sin\alpha,\ \
r'(h)|_{R\cos\alpha}=-\cot\alpha. \label{eq:bcrh}
\eq
Far away from the capsid, the membrane becomes flat. we have
\bq
h'(r)|_{\infty}=0\label{eq:bcinfhr}\eq or \bq
r'(h)|_{\infty}=\infty.\label{eq:bcinfrh}
\eq

Solving the elastic equation (\ref{eq:EL1}) with the boundary
conditions, Eq. (\ref{eq:bcrh}) and (\ref{eq:bcinfrh}) (or Eq.
(\ref{eq:bchr}) and Eq. (\ref{eq:bcinfhr}) if $h(r)$ is used), one
obtains the membrane shape that minimizes $\varepsilon_m$.
Substituting this shape into Eq.~(\ref{eq:Helfrich}), one obtains
the minimal $\varepsilon_m(\alpha)$. Putting its value into
Eq.~(\ref{eq:f}), one gets the total free energy density profile
$f(\alpha)$. In general, the elastic equation, Eq. (\ref{eq:EL1}),
is  highly non-linear and numerical calculations are needed to
obtain the exact membrane profile, as shown in
Sec.~\ref{sec:numerical}. However, in certain asymptotic limits,
analytical solutions can be obtained which determine the scaling
behavior of the system. This is done in the next two sections.

\section{Asymptotic solutions of the membrane elastic energy\label{sec:elastic}}

In calculating the free energy profile, the most nontrivial part
is to find the minimal $\varepsilon_m(\alpha)$, due to the
nonlinear elastic equation involved. After the solution is found,
it is straightforward to add the other part of the energy
$\varepsilon_c(\alpha)$ and get $f(\alpha)$. Therefore we focus on
the solution of minimal $\varepsilon_m$ in this section. Although
not solvable in general, the problem do have analytical solutions
in asymptotic limits. To a large extent, they determine the
analytical behavior of the system, especially the scaling behavior
of $\varepsilon_m$ with the dimensionless parameter \bq
\widetilde{\sigma}=R\sqrt{\frac{\sigma}{\kappa}},\eq which
characterize the relative strength of the surface tension to the
bending rigidity.

\subsection{The small deformation solution}

A typical approach to consider the elastic deformation of the
membrane is to take the small deformation approximation which
assumes $|\nabla h|\ll 1$~\cite{larged}. Here we use the notation
\bq
\nabla=\hat{r}\partial_r+\mathbf{\hat{\phi}}\frac{1}{r}\partial_\phi\eq
in order to show similarity of the elastic equation to the
linearized Poisson-Boltzmann equation later. Expanding with
$\nabla h$ and keeping terms of $O(\nabla h)^2$ in
$\delta\varepsilon_m=0$, we reach a linearized elastic equation
which can be written as \bq
H=\frac{1}{2}\nabla^2h,\nonumber\\
\nabla^2H-\frac{H}{r_s^2}=0,\label{eq:Hr}\eq where we have
introduced an important length scale in the problem, \bq
r_s=\sqrt{\frac{\kappa}{\sigma}}.\label{eq:r_s}\eq It is the
length scale beyond which the stretching energy becomes more
important than the bending energy. Notice that Eq.~(\ref{eq:Hr})
takes exactly the same form as a linearized Poisson-Boltzmann
equation in electrolytes or plasma~\cite{Landau-sm}. Therefore
$r_s$ can be interpreted as an elastic screening length, similar
to the Debye-H\"{u}ckel screening radius. The local curvature
$H(r)$ induced by the capsid decreases when $r$ increases and
becomes exponentially small at projected distance larger than
$r_s$. This is a typical linear solution of small deformation.

Using boundary conditions Eqs.~(\ref{eq:bchr}) and
(\ref{eq:bcinfhr}), the special solution to Eq.~(\ref{eq:Hr}) is
given by
\bq h(r)&=&
R\cos\alpha+r_s\tan\alpha\frac{K_0(r/r_s)-K_0(R\sin\alpha/r_s)}{
K_1(R\sin\alpha/r_s)},\label{eq:hrs}\nonumber\\
\\h'(r)&=&-\tan\alpha\frac{K_1(r/r_s)}{
K_1(R\sin\alpha/r_s)},\label{eq:h'r}\\H(r)&=&\frac{\tan\alpha}{2r_s}\frac{
K_0(r/r_s)}{K_1(R\sin\alpha/r_s)},\label{eq:Hrs}\eq
where $K_0$ and $K_1$ are the zero and first order modified Bessel
function of the second kind. At $r\gg r_s$, both $K_0(r/r_s)$ and
$K_1(r/r_s)$ decay like $\sqrt{r_s/r}\exp(-r/r_s)$, and the
deformation becomes exponentially small, as the meaning of $r_s$
suggested.

Substituting this solution back to Eq.~(\ref{eq:Helfrich}), we get the
minimal elastic energy of the membrane \bq
\varepsilon_m=\pi\kappa\tan^2\alpha\frac{R\sin\alpha}{r_s}
\frac{K_0(R\sin\alpha/r_s)}{K_1(R\sin\alpha/r_s)}.\label{eq:em}\eq
Notice that this energy is proportional to the dimensionless
parameter $ \widetilde{\sigma}=R\sqrt{\sigma/\kappa}=R/r_s.$ Here
the inverse proportion to $r_s$ is again a generic feature shared
with the theory of Debye-H\"{u}ckel linear
screening~\cite{Landau-sm}.

The self-consistency of the small deformation approximation is
warranted by $|h'(r)|<1$, or, according to Eq.~(\ref{eq:h'r}),
$|\tan\alpha|<1$. Therefore this solution is applicable in the
whole range of $r$ for $\alpha<\pi/4$ capsids only. On the other
hand, at large distances far away enough from the capsid, the deformation
of the membrane always becomes small enough such that the small
deformation solution is applicable. In this sense, this solution
can \emph{always} serve as a ``far-capsid" solution for the
membrane shape, although the formula for $\varepsilon_m$ in
Eq.~(\ref{eq:em}) is not valid in general. It describes the
universal decaying behavior of the deformation when the
deformation itself becomes small enough. We can formally define a
characteristic distance $r_c$ through \bq |h'(r_c)|=1,
\label{eq:hrc}\eq beyond which the small deformation solution is
valid. $r_c$ will be useful later when we discuss the complete
solution to the problem.

With the small deformation solution in hand, we are now ready to
derive a quantitative condition for the ideal capsid approximation
introduced in the last section. Clearly, when the average
projected distance between capsids, $d_0$, is much larger than
$r_s+2R$, the membrane mediated interaction between capsids is
negligible, since the deformations of the membrane by the capsids
at distance larger than $r_s$ are screened out. In this case, most
of the membrane surface is flat, so $d_0\simeq d$ (notice that $d$
is measured along the membrane surface which in general is larger
than $d_0$ measured along $r$ axis). Thus according to
Eqs.~(\ref{eq:cG}) and (\ref{eq:d}), the ideal capsids
approximation is valid when \bq
\frac{d_0}{r_s+2R}=\frac{2\sin(\alpha/2)}{(r_s+2R)\sqrt{\pi
c_G}\sin(\alpha_G/2)}\gg 1.\eq In this work, we assume $c_G$ is
small enough and this is always the case.

\subsection{The catenoid solution}

When the surface tension $\sigma=0$, or, $r_s\rightarrow\infty$,
again an analytical solution is available~\cite{kappa0}. In this
case, the second integral in Eq.~(\ref{eq:Helfrich}) is zero. Our
problem of finding the minimal $\varepsilon_m$ is reduced to a
minimal surface problem in differential geometry~\cite{Kreyszig}.
Namely, we look for the solution to the equation $H=0$~\cite{H}.
The only solution under the rotational symmetry of our problem is
the catenoid solution, first discovered by Euler in
1740~\cite{Morgan}.

\begin{figure}[ht]
\begin{center}
\includegraphics[width=0.5\textwidth]{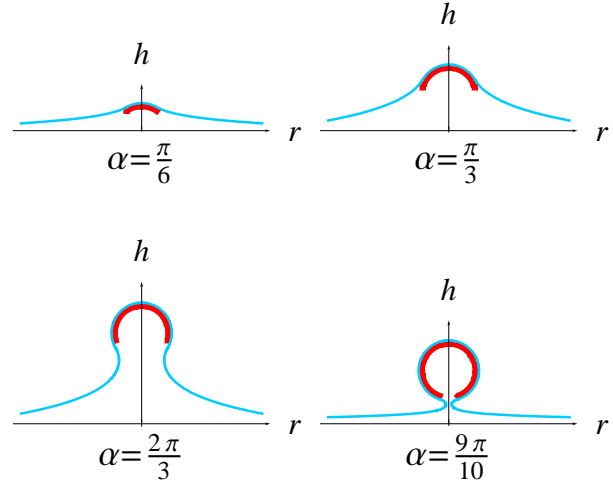}
\end{center}
\caption{(Color online). In the limit of small surface tension,
the optimal membrane shapes (thin line, blue online) around the
capsid (thick line, red online) are catenoids, as shown for
different capsid sizes.} \label{fig:cat}
\end{figure}

In this case, due to the possible multiple values of $h$ at the
same $r$, it is better to use the $r(h)$ representation. $H$ is
then given by Eq.~(\ref{eq:H}). Using boundary conditions
(\ref{eq:bcrh}) and (\ref{eq:bcinfrh}), the special solution to
$H=0$ is 
\bq 
&&r(h)=R\sin^2\alpha \nonumber \\
&&\quad\cosh
\frac{h-R\cos\alpha-R\sin^2\alpha\arcsinh(\cot\alpha)}{R\sin^2\alpha} .
\label{eq:rh}
\eq
The catenoid shapes for various
$\alpha$ are depicted in Fig.~~\ref{fig:cat}. In this catenoid
shape, the elastic energy $\varepsilon_m$ achieves its absolute
minimum, zero.

The catenoid solution is a solution to a nonlinear differential
equation. It involves large deformations which can not be
characterized by the linear solution discussed in the last
subsection. Although exact only when $r_s\rightarrow\infty$, this
solution is still useful for large but finite $r_s$~\cite{cat}. In
fact, since there are no other length scales in the elastic
equation (\ref{eq:EL1}) ($R$ only shows up in the boundary
conditions), a large $r_s$ actually means $r_s\gg r$. Therefore in
the region of $r\ll r_s$, the catenoid solution should work
asymptotically. In this sense, this solution can \emph{always}
serve as a ``near-capsid" solution for the membrane shape,
although $\varepsilon_m=0$ is not true in general. The
characteristic length beyond which it fails is simply $r_s$.

In case of $r_s\rightarrow\infty$ and $\alpha\ll 1$, both the
catenoid solution and the small deformation solution work in all
range of $r$. Indeed they become identical.

\subsection{Membrane elastic energy at two asymptotic limits}

The two solutions discussed in the last two subsections determine
the analytical behavior of the system to a large extent. When
$\widetilde{\sigma}\ll 1$, they can be combined to get the
analytical expression of the minimal $\varepsilon_m$. In general,
they determine the scaling behavior of $\varepsilon_m$ with
respect of $\widetilde{\sigma}$. We separate our discussion into
two opposite limits of small and large $\widetilde{\sigma}$, which
can be called the soft membrane regime and the stiff membrane
regime. Here ``soft" means easy to stretch, ``stiff" means the
opposite.

In the soft membrane regime, $\widetilde{\sigma}\ll 1$ or $R\ll
r_s$, the catenoid solution is valid near the capsid when $r\ll
r_s$. Calculating $r_c$ using Eqs.~(\ref{eq:rh}) and
(\ref{eq:hrc}), we get \bq r_c=\sqrt{2}R\sin^2\alpha.\eq  We see
$r_c\ll r_s$. Therefore the valid regions of the two asymptotic
solutions (one is $r>r_c$, the other is $r<r_s$) overlap largely
and we can combine them to get a complete solution to the optimal
membrane shape. Quantitatively, we artificially choose a projected
distance somewhere between $r_c$ and $r_s$, say $\sqrt{r_cr_s}$.
For $r<\sqrt{r_cr_s}$, the catenoid solution is used. For
$r>\sqrt{r_cr_s}$, the small deformation solution is used. Notice
that the special solution of the small deformation has to be
calculated using the continuity conditions for $h(r)$ and $h'(r)$
at $\sqrt{r_cr_s}$, derived from the catenoid solution. As a
result, we have an analytical expression for the optimal membrane
shape continuously from the edge of the capsid to infinity. The
corresponding $\varepsilon_m$, keeping the leading order terms in
the small parameter $\widetilde{\sigma}$, is given by
\bq
\varepsilon_m=\pi\kappa\sin^4\alpha\frac{R^2}{r_s^2}\ln\frac{r_s}{R}\label{eq:emb}.\eq
When $\alpha\ll 1$, this result agrees with the small deformation
solution in Eq.~(\ref{eq:em}) in the same regime of small
$\widetilde{\sigma}$. In this limit, we have $\varepsilon\propto
\widetilde{\sigma}^2\ln(1/\widetilde{\sigma})$.

In the stiff membrane regime, $\widetilde{\sigma}\gg 1$ or $R\gg
r_s$. Since $r\gg r_s$ always, the ``near capsid" region where the
catenoid solution holds disappears. On the other hand, for
$\alpha<\pi/4$, the small deformation solution is valid in the
whole range of $r$. The membrane elastic energy is given by
(\ref{eq:em}), which in this limit reads, \bq
\varepsilon_m=\pi\kappa\tan^2\alpha\sin\alpha\frac{R}{r_s}
.\label{eq:ema}\eq For $\alpha>\pi/4$ capsids, a rough estimate of
$r_c$ using Eq.~(\ref{eq:h'r}) gives \bq r_c\simeq
R\sin\alpha+r_s\ln|\tan\alpha|.\eq Since $R\gg r_s$, for most of
$\alpha$, we expect that the small deformation solution starts to
work at places close to the capsid. Probably because of this, the
scaling behavior of $\varepsilon_m\propto \widetilde{\sigma}$ is
preserved even at large $\alpha$, as shown by the numerical result
(see Sec.~\ref{sec:numerical}).

\section{Analytical result of the total free energy density\label{sec:analytical}}

After the information about the minimal $\varepsilon_m(\alpha)$ is
known, we can add the line tension energy $\varepsilon_c(\alpha)$
to it and consider the total free energy density $f(\alpha)$. The
presence of $\varepsilon_c$ introduces the second dimensionless
parameter to the problem, \bq
\widetilde{\tau}=\frac{R\tau}{\kappa},\eq which characterize the
relative strength of the line tension on the capsid rim. In this
section, we derive several simple scaling behaviors of the system,
depending on the two dimensionless parameters $\widetilde{\sigma}$
and $\widetilde{\tau}$. We again separate our discussion into the
soft and stiff membrane regimes corresponding to small and large
$\widetilde{\sigma}$.

\subsection{The soft membrane regime}

In the soft membrane regime, $\widetilde{\sigma}=R/r_s\ll 1$.
Substituting Eqs.~(\ref{eq:Ec}) and (\ref{eq:em}) to Eq.~(\ref{eq:f}),
we have \bq f&\equiv&\kappa
c_G(1-\cos\alpha_G)\widetilde{f}\nonumber\\&=&\kappa
c_G(1-\cos\alpha_G)\pi\cot\frac{\alpha}{2}(2\widetilde{\tau}+
\widetilde{\sigma}^2\ln\frac{1}{\widetilde{\sigma}}
\sin^3\alpha),\nonumber\\\label{eq:fcat}\eq where we have
introduced the dimensionless free energy density $\widetilde{f}$
for convenience. $\widetilde{f}(\alpha)$ is plotted schematically
in Fig.~\ref{fig:free}. When $\widetilde{\tau}$ is large, the only
minimum of the free energy density is at $\alpha\rightarrow \pi$
(the left panel of Fig.~\ref{fig:free}). On the other hand, when
$\widetilde{\tau}<
0.2\widetilde{\sigma}^2\ln(1/\widetilde{\sigma})$, a local minimum
at the capsid size, $\alpha_0$, appears ( the right panel of
Fig.~\ref{fig:free}). Correspondingly, the threshold line tension
at which the local minimum in the free energy density appears is
\bq\tau_c=0.2R\sigma\ln\frac{1}{R\sqrt{\sigma/\kappa}}.\label{eq:tca}\eq

Since transcendental equations are involved in minimization of
$\widetilde{f}$, it is not easy to get the analytical expression
about this local minimum in general. However, $\alpha_0$ and the
corresponding $\widetilde{f}_0$ can be estimated in a linear
approximation. Assuming $\alpha_0$ is achieved at small $\alpha$,
we can expand $\widetilde{f}$ and keep only the leading order
terms in $\alpha$. We get \bq
\widetilde{f}=\frac{4\pi\widetilde{\tau}}{\alpha}
+2\pi\widetilde{\sigma}^2\ln\frac{1}{\widetilde{\sigma}}\alpha^3.
\eq Taking $\partial f/\partial\alpha=0$, we have \bq\alpha_0=
\sqrt[\uproot{15}3]{\frac{\widetilde{\tau}}
{\widetilde{\sigma}^2\ln(1/\widetilde{\sigma})}}
=\sqrt[\uproot{15}3]{\frac{\tau}{R\sigma\ln(\sqrt{\kappa/\sigma}/R)}}.\label{eq:alpha0s}\eq
The fact that this is a minimum rather than a maximum is confirmed
by $\partial^2 \widetilde{f}/\partial\alpha^2|_{\alpha_0}>0$. For
$\tau<\tau_c$ at which $\alpha_0$ shows up, this result is indeed
much smaller than one, consistent with the initial assumption that
$\alpha_0\ll 1$. In the same limit,
\bq\widetilde{f}_0\simeq 4\pi\sqrt[\uproot{4}3]{\widetilde{\tau}^2
\widetilde{\sigma}^2\ln(1/\widetilde{\sigma})}
=4\pi\frac{R}{\kappa}\sqrt[\uproot{15}3]{\tau^2\sigma\ln\frac{\sqrt{\kappa/\sigma}}{R}}.\label{eq:f0s}\eq

\subsection{The stiff membrane regime}

In this case, we do not know the form of the membrane elastic
energy $\varepsilon_m$ for large $\alpha$. Still, in the same
spirit of linear analysis, we can assume that there is a minimum
of $f$ at small $\alpha$, and use the small deformation solution
Eq.~(\ref{eq:ema}) for $\varepsilon_m$. Notice that the minimum
found in this way is only a local minimum, since we did not
include the information of large $\alpha$.

As a result, we have \bq \widetilde{f}
&=&\pi\cot\frac{\alpha}{2}(2\widetilde{\tau}+
\widetilde{\sigma}\tan^2\alpha)\nonumber\\
&\simeq&\frac{4\pi\widetilde{\tau}}{\alpha}+2\pi\widetilde{\sigma}\alpha.\eq
Taking $\partial f/\partial\alpha=0$, we get
\bq\alpha_0=\sqrt{\frac{2\widetilde{\tau}}{\widetilde{\sigma}}}
=\sqrt[\uproot{10}4]{\frac{4\tau^2}{\kappa\sigma}}.\label{eq:alpha0L}\eq
It is a minimum since $\partial^2
\widetilde{f}/\partial\alpha^2|_{\alpha_0}>0$. For this result to
be meaningful, $\widetilde{\tau}\ll\widetilde{\sigma}$ must hold,
which will be checked in comparison with the numerical result. The
corresponding free energy density is \bq
\widetilde{f}_0=4\pi\sqrt{2\widetilde{\sigma}\widetilde{\tau}}
=4\pi
R\sqrt[\uproot{15}4]{\frac{4\sigma\tau^2}{\kappa^3}}.\label{eq:f0L}\eq

\section{Numerical result and Discussion\label{sec:numerical}}

In order to verify our analytical understanding and get the
complete solution to the problem, we solve the nonlinear elastic
equation derived from $\delta\varepsilon_m=0$ numerically. Our
computation procedure follows Refs.~\cite{Deserno04,Lipowsky91}.
This numerical solution is then combined with $\varepsilon_c$ to
give the total free energy density $f$. In this section, we show
the numerical result, compare it with the analytical formulas, and
discuss the meaning of our results.

\begin{figure}[h]
\begin{center}
\includegraphics[width=0.51\textwidth]{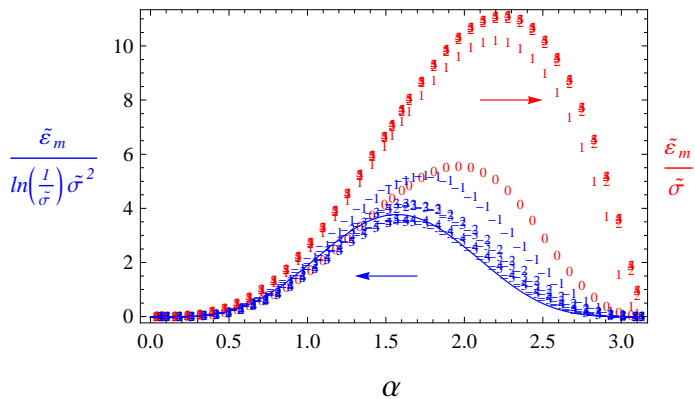}
\end{center}
\caption{(Color online) Numerical result of the dimensionless
membrane elastic energy
$\widetilde{\varepsilon}_m=\varepsilon_m/\kappa$ as a function of
$\alpha$. The eleven sets of data points are at
$\widetilde{\sigma}=10^{-5},10^{-4},10^{-3},...,10^{5}$. They are
labelled correspondingly as $-5,-4,-3,...5$. The left axis
$\widetilde{\varepsilon}_m/\widetilde{\sigma}$ is for all
$\widetilde{\sigma}\geq 1$ data points (red on line), while the
right axis
$\widetilde{\varepsilon}_m/\widetilde{\sigma}^2\ln(1/\widetilde{\sigma})$
is for all $\widetilde{\sigma}<1$ data points (blue on line), as
indicated by the two arrows. The curve represents the analytical
asymptotic solution (\ref{eq:emb}) with an additional factor $1.3$
(blue on line), fitting the data points for $\widetilde{\sigma}\ll
1$.} \label{fig:em}
\end{figure}

The direct numerical result of $\varepsilon_m$ is plotted in
Fig.~\ref{fig:em}, where for convenience we used the dimensionless
elastic energy $\widetilde{\varepsilon}_m=\varepsilon_m/\kappa$.
The first important thing to notice is that the elastic energy
profile always takes a ``sand dune" shape, where two minimums,
zeros, are achieved at $\alpha\rightarrow 0,\pi$, and a maximum
shows up in the middle of $\alpha$. Physically, this energy
profile comes from the need of matching boundary conditions at the
edge of the capsid and at infinity. The membrane deformed by the
capsid edge at one end has to become flat far away from the
capsid. At $\alpha\rightarrow 0$ and $\alpha\rightarrow\pi$, the
membrane is not deformed at all, and the elastic energy is
zero~\cite{neck}. While for $\alpha$ close to $\pi/2$, the
membrane is almost vertical at the edge of the capsid, and a large
amount of elastic energy is needed to bend it flat.

Secondly, we see clearly two kinds of asymptotic behaviors of
$\varepsilon_m$ depending on the parameter
$\widetilde{\sigma}=R/r_s$. In the stiff membrane regime,
$\widetilde{\sigma}\gg 1$, the energy is proportional to
$\widetilde{\sigma}$ as shown by the collapse of the data points
to a single curve with $\widetilde{\sigma}$ varying from $10^2$ to
$10^5$. The maximum of the energy is achieved at $\alpha_m\simeq
0.7\pi$. $\alpha_m$ is a nonlinear result and can not be
calculated analytically. However, the proportionality of
$\varepsilon_m$ to $\widetilde{\sigma}$ is a small deformation
result as shown in Eq.~(\ref{eq:ema}). In the soft membrane
regime, $\widetilde{\sigma}\ll 1$, the energy is proportional to
$\widetilde{\sigma}^2\ln(1/\widetilde{\sigma})$, shown again by
the collapse of the data points with $\widetilde{\sigma}$ varying
from $10^{-2}$ to $10^{-5}$. Here the collapse is not as
pronouncing as in the other regime mostly due to the larger
numerical error in dealing with smaller
$\widetilde{\varepsilon}_m$. The absolute value of
$\widetilde{\varepsilon}_m$ in this regime is smaller at least in
four order of magnitude than in the other regime. The maximum of
the energy here is arrived at $\alpha_m=\pi/2$ and the curve
becomes symmetric about $\alpha_m$. These features agree with our
small $\widetilde{\sigma}$ solution originating from the catenoid
solution. In fact, Eq.~(\ref{eq:emb}) fits the numerical data
reasonably well, with an additional factor $1.3$.

\begin{figure}[h]
\begin{center}
\includegraphics[width=0.4\textwidth]{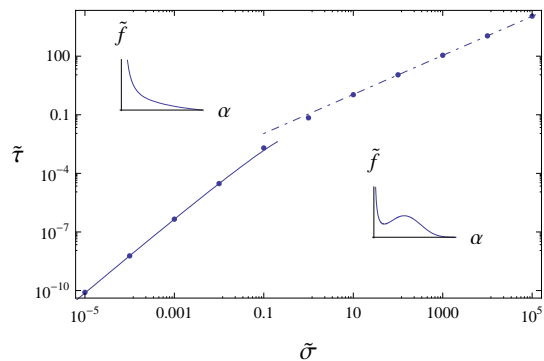}
\end{center}
\caption{An effective ``phase diagram" in the plane of two
dimensionless parameters, $\widetilde{\sigma}=R/r_s$ and
$\widetilde{\tau}=R\tau/\kappa$. In the upper-left part, the free
energy density decreases monotonically with $\alpha$, while in the
lower-right part, it has a local minimum, as shown in the insets.
The numerical data points mark the ``phase boundary" at which the
local minimum appears. The dotdashed line and the solid lines fit
the data points using $\widetilde{\tau}=0.11 \widetilde{\sigma}$
and
$\widetilde{\tau}=0.065\widetilde{\sigma}^2\ln(1/\widetilde{\sigma})$
respectively.} \label{fig:phase}
\end{figure}

The scaling of $\varepsilon_m$ with $\widetilde{\sigma}$ suggests
a simple way to do the numerical calculation to the free energy
density. When $\widetilde{\sigma}\gg 1$,
\bq
\widetilde{f}=\widetilde{\varepsilon}_c+\widetilde{\varepsilon}_m=2\pi\cot\frac{\alpha}{2}\widetilde{\tau}
+g_1(\alpha)\widetilde{\sigma}\nonumber\\
=\widetilde{\sigma}\left[2\pi\cot\frac{\alpha}{2}\frac{\widetilde{\tau}}{\widetilde{\sigma}}
+g_1(\alpha)\right],\eq
where $\widetilde{\varepsilon}_c=\varepsilon_c/\kappa$ and
$g_1(\alpha)$ is some function given by the numerical computation.
According to the last equality, up to an overall constant
$\widetilde{\sigma}$, $\widetilde{f}$ is completely determined by
only one parameter $\widetilde{\tau}/\widetilde{\sigma}$.
Similarly, when $\widetilde{\sigma}\ll 1$, \bq
\widetilde{f}=\widetilde{\sigma}^2 \ln\frac{1}{\widetilde{\sigma}}
\left[2\pi\cot\frac{\alpha}{2}\frac{\widetilde{\tau}}{\widetilde{\sigma}^2
\ln(1/\widetilde{\sigma})}+g_2(\alpha)\right],\eq where
$g_2(\alpha)$ is again given by numerical computation, although we
know it from our analytical result Eq.~(\ref{eq:fcat}). In this
regime, $\widetilde{f}$ is determined by one parameter
$\widetilde{\tau}/\widetilde{\sigma}^2 \ln(1/\widetilde{\sigma})$.
Below in studying the local minimum of $\widetilde{f}$, we
therefore consider a single parameter dependence.

\begin{figure}[h]
\begin{center}
\includegraphics[width=0.45\textwidth]{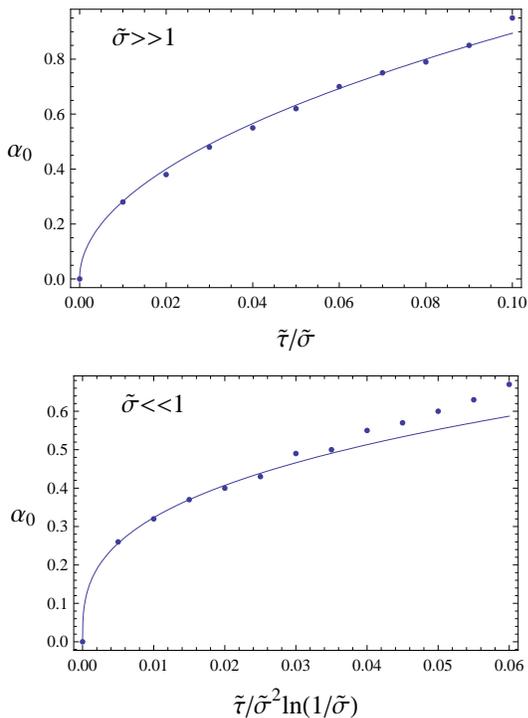}
\end{center}
\caption{The capsid size $\alpha_0$ as a free energy local minimum
is shown as a function of $\widetilde{\tau}/\widetilde{\sigma}$
and $\widetilde{\tau}/\widetilde{\sigma}^2
\ln(1/\widetilde{\sigma})$ at two limits of $\widetilde{\sigma}$.
The dots are numerical results taken at $\widetilde{\sigma}=10^2,
10^3, 10^4, 10^5$ for the upper panel and
$\widetilde{\sigma}=10^{-2}, 10^{-3}, 10^{-4}, 10^{-5}$ for the
lower panel. The curves are analytical results of
Eq.~(\ref{eq:alpha0L}) at $\widetilde{\sigma}\gg 1$ and
Eq.~(\ref{eq:alpha0s}) at $\widetilde{\sigma}\ll 1$ with
additional numerical factors of 2 and $1.5$ respectively. The
range of $\widetilde{\tau}$ plotted corresponds to the lower-right
``phase" in the ``phase diagram" of Fig.~\ref{fig:phase}.}
\label{fig:minim}
\end{figure}

For all $\widetilde{\sigma}$ and $\widetilde{\tau}$, we get two
different types of free energy density profiles as shown in
Fig.~\ref{fig:free}, consistent with the analytical result for
small $\widetilde{\sigma}$. The global minimum of the free energy
density is always at $\alpha\rightarrow\pi$. Physically, the line
tension energy prefers the shortest length of the capsid rim,
which is zero for complete capsids ($\alpha\rightarrow\pi$). When
$\widetilde{\tau}$ is very large, the line tension energy
dominates, and the free energy density $\widetilde{f}$ decreases
with $\alpha$ monotonically to zero, as shown in the left panel of
Fig.~\ref{fig:free}. On the other hand, when $\widetilde{\tau}$ is
small, due to the maximum of the membrane elastic energy
$\varepsilon_m$, a local minimum at the capsid size, $\alpha_0$,
shows up in the free energy density, as shown in the right panel
of Fig.~\ref{fig:free}. It is useful to draw a ``phase diagram" on
the plane of $\widetilde{\sigma}$ and $\widetilde{\tau}$ as
Fig.~\ref{fig:phase} to show this qualitative difference in the
free energy density profile. 
The lower right region of Fig. \ref{fig:phase} corresponds to
value of the parameters ($\widetilde{\sigma}$, $\widetilde{\tau}$)
where capsid budding can be kinetically trapped.
The two lines fit the ``phase boundary" at
large and small $\widetilde{\sigma}$ with $ \widetilde{\tau}=0.11
\widetilde{\sigma}$ and
$\widetilde{\tau}=0.065\widetilde{\sigma}^2\ln(1/\sigma)$
respectively. As one can see, there is a very good agreement between
numberical results and our scaling formulas for $\widetilde{\sigma}$
in two asymptotic limits. According to the numerical
fits, the threshold $\tau$ at which the local minimum in
the free energy density shows up are
\bq \tau_c=0.11\sqrt{\kappa\sigma}\label{eq:tcn}\eq
when $\widetilde{\sigma}\gg 1$, and
\bq \tau_c=0.065 R\sigma\ln\frac{1}{R\sqrt{\sigma/\kappa}}\eq
when $\widetilde{\sigma}\ll 1$. The later formula agrees with our
analytical result Eq.~(\ref{eq:tca}) with a numerical factor 3
difference.

\begin{figure}[h]
\begin{center}
\includegraphics[width=0.45\textwidth]{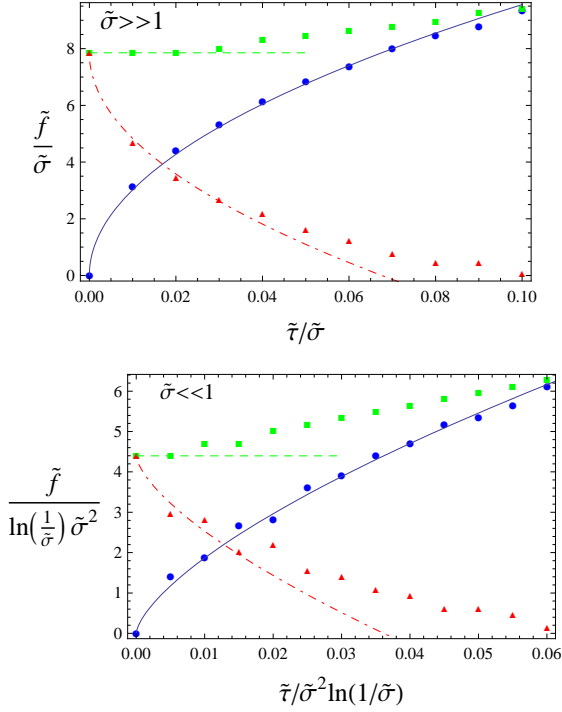}
\end{center}
\caption{(Color online). The local minimum $\widetilde{f}_0$,
maximum $\widetilde{f}_m$, and barrier
$\widetilde{f}_m-\widetilde{f}_0$ of the free energy density. The
values of $\widetilde{\sigma}$ and $\widetilde{\tau}$ plotted are
the same as in Fig.~\ref{fig:minim}. The circles (blue online) are
the numerical result of $\widetilde{f}_0$, fitted by the solid
lines (blue online) using Eq.~(\ref{eq:f0L}) at $\sigma\gg 1$ and
Eq.~(\ref{eq:f0s}) at $\sigma\ll 1$ with additional numerical
factor $1.7$ and $3.2$ respectively. The squares (green online)
are the numerical result of $\widetilde{f}_m$, marked by the
dashed lines (green online) at their zero $\widetilde{\tau}$
values. The triangles (red online) are numerical result of the
barriers, $\widetilde{f}_m-\widetilde{f}_0$, fitted by the
dotdashed lines (red online) using Eqs.~(\ref{eq:barL}) and
(\ref{eq:bars}).} \label{fig:barr}
\end{figure}

The possible local minimum in $\widetilde{f}$ (the right panel in
Fig.~\ref{fig:free}) suggests that budding may be trapped
kinetically at the capsid size $\alpha_0$. Numerical and
analytical results of $\alpha_0$ are shown in Fig~\ref{fig:minim}.
The analytical curves are drawn using Eq.~(\ref{eq:alpha0L}) at
$\widetilde{\sigma}\gg 1$ and Eq.~(\ref{eq:alpha0s}) at
$\widetilde{\sigma}\ll 1$, with additional numerical factors of 2
and $1.5$ respectively. There is a large deviation between
analytical and numerical results at large
$\widetilde{\tau}$. This is the parameter regime where the 
linear approximation is no longer valid.

The kinetic trapping becomes significant if the barrier in the
free energy density is large. In Fig.~\ref{fig:barr}, numerical
and analytical results about this barrier are plotted. For the
local minimum $\widetilde{f}_0$, up to a order one numerical
factor ($1.7$ and $3.2$), our analytical expressions
Eq.~(\ref{eq:f0L}) at $\sigma\gg 1$ and Eq.~(\ref{eq:f0s}) at
$\sigma\ll 1$ remains reasonable approximation. We can not estimate the maximum
$\widetilde{f}_m$, which is in the nonlinear regime. However, in
the most important regime of small $\widetilde{\tau}$ and large
barrier, Fig.~\ref{fig:barr} shows that the main contribution to
$\widetilde{f}_m$ comes from the membrane elastic energy
$\varepsilon_m$ (the value of $\widetilde{f}_m$ at
$\widetilde{\tau}=0$). In this regime, the additional contribution
to $\widetilde{f}_m$ from the line tension energy $\varepsilon_c$
is negligible and $\widetilde{f}_m$ is almost a constant.
Combining the numerical result of $\widetilde{f}_m$ and the
analytical result of $\widetilde{f}_0$ with proper numerical
factors, we get the asymptotic formulas for the barrier at
$\widetilde{\tau}\ll 1$,
\bq \widetilde{f}_m-\widetilde{f}_0&\simeq&7.8\widetilde{\sigma}
-6.8\pi\sqrt{2\widetilde{\sigma}\widetilde{\tau}} \nonumber\\
&=&7.8R\sqrt{\frac{\sigma}{\kappa}}-6.8\pi
R\sqrt[\uproot{15}4]{\frac{4\sigma\tau^2}{\kappa^3}},\ \
(\widetilde{\sigma}\gg 1).\label{eq:barL}\\
\widetilde{f}_m-\widetilde{f}_0&\simeq&4.4\widetilde{\sigma}^2\ln(1/\widetilde{\sigma})-
12.8\pi\sqrt[\uproot{4}3]{\widetilde{\tau}^2
\widetilde{\sigma}^2\ln(1/\widetilde{\sigma})}\nonumber\\
&=&4.4R^2\frac{\sigma}{\kappa}\ln\frac{\sqrt{\kappa/\sigma}}{R}-12.8\pi\frac{R}{\kappa}\sqrt[\uproot{15}3]{\tau^2\sigma
\ln\frac{\sqrt{\kappa/\sigma}}{R}},\nonumber\\ &&
(\widetilde{\sigma}\ll 1).\label{eq:bars} \eq The largest barriers
are achieved at $\widetilde{\tau}=0$ or $\widetilde{f}_0=0$.

\section{kinetics of budding and partial budding\label{sec:exp}}

As discussed in the previous sections, with a finite Gag-Gag
attraction, budding always proceeds to completion
thermodynamically. However, when this attraction is weak, or
$\tau$ is small, a metastable state of partial budding appears at
a smaller capsid size $\alpha_0$ (see Fig.~\ref{fig:free}). It is
therefore possible that the budding process is kinetically trapped
at $\alpha_0$. In this section, we discuss this kinetic effect and
make connections of our theory to experiments.

Let us first estimate the values of parameters. A normal plasma
membrane has $\kappa\simeq 20-40\mathrm{k_BT}$ and $\sigma\simeq
0.5-2 \mathrm{pN}/\mathrm{nm}=0.12-0.48
\mathrm{k_BT}/\mathrm{nm}^2$~\cite{Morris}. On the other hand,
typical HIV have $R\simeq 60 - 80\mathrm{nm}$~\cite{Coffin}.
Consequently, $\widetilde{\sigma}=R\sqrt{\sigma/\kappa}\simeq 10$
and only the stiff membrane regime with $\widetilde{\sigma}\gg 1$
is relevant for HIV. Actually in the opposite regime of
$\widetilde{\sigma}\ll 1$, one can show that the typical energy
scale
$\varepsilon_m\sim\kappa\widetilde{\sigma}^2\ln(1/\widetilde{\sigma})$
is comparable to $\mathrm{k_BT}$ and thus not important in the
room temperature. In this section, we therefore focus on the stiff
membrane regime only.

In order to see if budding can be kinetically trapped at the local
minimum $\alpha_0$ (see Fig.~\ref{fig:free}), we have to study the
budding kinetics and calculate the kinetic barrier. For this
purpose, let us employ the standard kinetic picture of the first
order phase transition~\cite{LandauKinetics,NguyenKinetics}, corresponding to the
transition from a free-Gags phase to an aggregated Gags phase
where Gags self-assemble into complete viral capsids. At the
initial stage of aggregation, the concentration of free Gags is
large, Gags coagulate to form dimers. Dimers coagulate with free
Gags or other dimers to form larger Gag clusters (small capsids).
This initial coagulation or nucleation is a fast process and is
not a rate limiting step in retroviral budding. Soon free Gags are
significantly depleted, and the main kinetic pathway for growth of
capsids is for them to diffuse and merge with each other. We will
be concern with this later stage of coagulation. For simplicity,
we work with the dominant capsid size, $\alpha(t)$ [with
concentration $n(t)$], assuming these typical capsids carry all
the mass of membrane-bound Gag proteins.

Let us start with the case when $\alpha(t)$ is still small so that
the energy barrier for merging of capsids is smaller than
$\mathrm{k_BT}$. This is the regime of the well known diffusion
limited aggregation~\cite{ColloidalDomain}. The rate of the capsid
area $A(\alpha,t)$ incretion is proportional to the probability
that two capsids diffuse and merge with each other. The kinetic
rate equation reads
\bq \frac{d A(\alpha,t)}{d t} = 2\pi R \sin\alpha(t) A(t) D \nabla
n(t)|_{R\sin\alpha}, \label{eq:dAdt}\eq
where $D \simeq k_BT/\eta R\sin\alpha(t)$ is the lateral diffusion
constant of a capsid on the membrane and $\nabla
n(t)|_{R\sin\alpha}$ is the gradient of the concentration $n(t)$
on the edge of the capsid. This gradient can be estimated assuming
a steady state in the diffusion and taking the adsorbing boundary
condition at the edge of the capsid and a given capsid
concentration [Eq.~(\ref{eq:cG})] far away from the capsid.
Solving the diffusion equation with these boundary conditions, we
find
\bq \nabla
n(t)|_{R\sin\alpha}=\frac{c_GA(\alpha_G)}{A(\alpha)R\sin\alpha}.\eq
Substituting these relations and Eq.~(\ref{eq:A}) into
Eq.~(\ref{eq:dAdt}), we get \bq
\frac{\alpha(t)}{2}-\frac{\sin2\alpha(t)}{4}=\frac{t}{\tau_{diff}}
+\frac{\alpha_G}{2}-\frac{\sin2\alpha_G}{4}. \eq
where
\bq
\tau_{diff}=\eta R^3/Tc_GA(\alpha_G)
\label{tauDiff}
\eq
is the time scale of diffusion proportional to the
viscosity $\eta$ of the membrane. In
the small $\alpha$ regime corresponding to a small kinetic
barrier, this equation can be written as
\bq\alpha(t)=(3t/\tau_{diff}+\alpha_G^3)^{1/3}, \eq
which is a slow function of time.

The regime of diffusion limited growth stops when the kinetic
barrier between approaching partial capsid is much larger than
$k_BT$. At a later time, a different growth regime of
Lifshitz-Slezov (LS) comes into play~\cite{LandauKinetics}. In
this mechanism, the growth is no longer due to collision and
merging of partially budded capsids. Instead, smaller capsids shrink and release
individual Gags. These Gag are absorbed into larger capsids,
leading to their growth. This process of releasing and adsorbing
of individual Gags (the so-called coalescence) has much smaller
kinetic barrier than the barrier to capsid merging in this later
stage. The growth of capsid size in LS regime is the same as
that of diffusion limited growth \cite{LandauKinetics}.
However, the rate constant $\tau_{LS}$ depends exponentially
on the activation energy to release individual Gag proteins
from a capsid
\bq
\tau_{LS} \propto \exp(-|\epsilon|/k_BT)
\eq
where $\epsilon$ is the binding energy of Gag in a capsid, which
itself is also a function of the Gag-Gag interaction. 

The kinetic picture described above is good when $\tau>\tau_c$ and
the free energy density decreases monotonically with increasing
$\alpha$ (the left panel of Fig.~\ref{fig:free}). However, when
$\tau<\tau_c$ and a local minimum $\alpha_0$ appears in the free
energy density (the right panel of Fig.~\ref{fig:free}), the above
picture must be modified. For the cluster growth, either
in the diffusion-limited regime or in the LS regime, the growth of
the cluster size always reduces the free energy of the system. On
the other hand, for the capsid growth of retroviral budding, after
the capsid size reaches $\alpha_0$, the system free energy
increases when the capsids grow further. For $\alpha>\alpha_0$,
the growth of capsids is determined by the ability to tunnel
through the kinetic barrier related to $f_m-f_0$ (see
Fig.~\ref{fig:free}). The detail analysis of the rate of capsid
growth for $\alpha>\alpha_0$ is a very interesting problem by
itself, requiring understanding of membrane energetics when a partially
budded capsid absorbs other capsids or many individual Gags to increase
its size from $\alpha_0$ to $\alpha_m$. These calculations are beyond the
scope of this paper and we will leave the detail
treament of capsid growth in this case to a future study. Nevertheless,
one can expect the rate of such process to inversely proportional
to the exponential of the energy barrier
\bq \tau_{tunnel} \propto \exp[-(f_m-f_0)/nk_BT], \eq
where $(f_m-f_0)/n$ is the energy barrier of a membrane cell
with a single capsid in it. According to Eqs.~(\ref{eq:barL}) (see
also the upper panel of Fig.~\ref{fig:barr}), the maximum energy
barrier is achieved at $f_0=0$ or $\widetilde{\tau}=0$. Using
Eqs.~(\ref{eq:cG}) and~(\ref{eq:fcat}), it can be written as
\bq E_m=\frac{f_m}{n}=\kappa(1-\cos\alpha_m)\widetilde{f}_m,\eq
where $\widetilde{f}_m$ is given by Eqs.~(\ref{eq:barL}), and
$\alpha_m$ is the corresponding capsid size. A more useful
expression of $E_m$ can be got if one recognizes that $E_m$ is
nothing but the maximum of $\varepsilon_m$ shown in
Fig.~\ref{fig:em}. Using the numerical result of that figure, we
get
\bq
E_m=11.5\kappa\widetilde{\sigma}=11.5R\sqrt{\kappa\sigma}.\label{eq:Em}
\eq
Clearly, $E_m\gg \mathrm{k_BT}$ for $\widetilde{\sigma}>1$. For
example, for $R=70 \texttt{nm}$, $\sigma=0.24 \mathrm{k_BT/nm^2}$
and $\kappa=20 \mathrm{k_BT}$, we get $E_m=1765 \mathrm{k_BT}$.
The true energy barrier is smaller than $E_m$ since
$\widetilde{\tau}> 0$. In the regime of small $\widetilde{\tau}$,
according to Eq.~(\ref{eq:barL}), it is
\bq E&\simeq&\kappa(1-\cos\alpha_m)
(\widetilde{f}_m-\widetilde{f}_0)\nonumber\\
&=&E_m\left(1-3.9\sqrt{\frac{\widetilde{\tau}}{\widetilde{\sigma}}}\right)
=E_m\left(1-3.9\sqrt[\uproot{10}4]{\frac{\tau^2}{\kappa\sigma}}\right).\label{eq:E}
\eq

In experiments, if one treats $\kappa$ and $\sigma$ as constants
then according to Eq.~(\ref{eq:Em})
the larger the retrovirus size $R$, the bigger the kinetic
barrier. On the other hand, the line tension $\tau$ is directly
proportional to the strength of the Gag-Gag attraction and is
experimentally adjustable through mutation of the late domain on
the Gag protein, binding of other molecules to Gags or changing
the pH, salinity of water solution near the membrane \cite{Sundquist04,ReinPNAS01}. 
As we know, the closest approach distance between two Gag proteins
is about $10\mathrm{nm}$~\cite{Coffin}. If Gags are densely packed on
the capsid, this gives $\tau\simeq
0.5\mathrm{k_BT}/\mathrm{nm}\simeq 2 \mathrm{pN}$ for normal
retroviral capsids. Theoretically, in order to have a local
minimum in the free energy density and trap retrovirus budding
kinetically, we must have $\tau<\tau_c$ (see Fig.~\ref{fig:free}).
For a normal cell membrane with $\kappa=20 \mathrm{k_BT}$ and
$\sigma=0.24 \mathrm{k_BT}/\mathrm{nm}^2$, using
Eq.~(\ref{eq:tcn}), $\tau_c\simeq
0.24\mathrm{k_BT}/\mathrm{nm}=1\mathrm{pN}$. Therefore for normal
capsids, $\tau>\tau_c$, and budding easily proceeds to completion
(see the left panel of Fig.~\ref{fig:free}). One the other hand,
$\tau$ is bigger than $\tau_c$ only by a factor of 2. Therefore
HIV budding can be fairly easily trapped at a partially budded
state with capsid size $\alpha_0$ by reducing the Gag-Gag
interaction strength such as mutation of a single domain on the
Gag protein. The kinetic barrier $E$ appeared at $\alpha=\alpha_0$
can be much larger than $\mathrm{k_BT}$, and the time scale for
capsid growth beyond $\alpha_0$, $\tau_{tunnel}$, is exponentially
large. Qualitatively, this trend is consistent with experiments on
mutation of the late domain of Gag
proteins~\cite{Sundquist04,Freed04}. Numerically, we know that
$\alpha_0<0.9\simeq 0.3\pi$ (see the upper panel of
Fig.~\ref{fig:minim}). It agrees with experiments reasonably
well, although there are many additional factors that we neglected
in our treatment such as local variation in membrane elasticity
due to raft structures or the presence of other proteins in
in-vivo assembly and budding. More controlled experiments are needed to
to verify the dependence on the membrane rigidities and Gag-Gag
attraction of $\alpha_0$ given by Eq.~(\ref{eq:alpha0L}).

\section{Conclusion\label{sec:conclusion}}

In this paper, we developed an elastic model of HIV (and
retroviruses in general) budding and self-assembly on the elastic
membrane. We studied the free energy profile of the system as a
function of the capsid size $\alpha$. We showed that although
always thermodynamically favorable, complete budding and assembly
may not be achieved if the Gag-Gag attraction is weak. In practice,
for normal biological conditions, the Gag-Gag attraction is strong
enough and HIV budding and assembly always proceed to completion,
as it should be. On the other hand, it is fairly easy to trap HIV
budding to a partially budded state with capsid size $\alpha_0$ by
reducing the Gag-Gag attraction. This can be done
through the mutation of late domain on the Gag protein or binding
of other molecules to Gag. In principle, the trapping is also possible by
increasing the membrane rigidities, although this is not easy to
do in vivo. Our theory agrees with reasonably well with experimental
results. However, experiments with better controlled environments are
 needed to verify various aspect of the theory.

The most interesting point of our
model is probably that it provides a unique self-assembly
mechanism. Not like self-assembly of other viruses or colloids,
HIV assemble and bud \emph{concurrently} on the membrane.
Therefore the membrane elastic energy plays an important role in
the assembly process. For example, the kinetic barrier which traps
the HIV budding essentially comes from the membrane elastic energy
around the capsids. In fact, our model developed for HIV budding and assembly can be
very well applied to other situations. For example, for a given
concentration of membrane-bounded proteins with a fixed
spontaneous curvature, this kind of budding and assembly
phenomenon should also exist and can be explained using our model.
In this situation, it may be easier to change the membrane
properties and protein-protein attraction in vitro to verify our
theory more quantitatively. Due to the interplay between the 
membrane elastic energy and the Gag-Gag attraction energy, 
the kinetics of retrovirus budding is an interesting problem
by itself, as discussed in Sec.~\ref{sec:exp}. We plan to 
address this question in more detail the near future.

\begin{acknowledgments}
We wish to thank G. Bel, J. Mueller, B. I. Shklovskii and T. A. Witten for
useful discussions. T.T.N. acknowledges the junior faculty
support from the Georgia Institute of Technology.
\end{acknowledgments}

\bibliography{buddingbib}

\begin{thebibliography}{32}
\bibitem[{G()}]{G}
\bibinfo{note}{In Ref.~\cite{Lipowsky96}, the Gaussian curvature term is
  important since the area of the budding region is not fixed.}
\bibitem[{lar()}]{larged}
\bibinfo{note}{Expansion in the opposite limit $|\nabla h|\gg 1$ gives a
  nonlinear differential equation which can not be solved analytically.}
\bibitem[{kap()}]{kappa0}
\bibinfo{note}{In the opposite limit of $\kappa=0$, although
  $\delta\varepsilon_m=0$ still has a catenoid solution, it is only a
  stationary solution but does not correspond to an energy minimum. Actually
  $\varepsilon_m$ can be arbitarily close to zero but not equal to zero, given
  a membrane shape arbitarilly close to the flat membrane and only deformed a
  little bit at the capsid rim. This result is different from the minimal
  surface of revolution problem in the calculus of variation~\cite{Brunt}. This
  is eccentially due to the fact that our boundary condition requires the
  membrane to be flat at infinity, but there is no confinment to its position
  there.}
\bibitem[{H()}]{H}
\bibinfo{note}{Since all energies we considered are positive definite, $H=0$
  corresponds an absolute minimum for $\varepsilon_m$. One can of course still
  use $\delta \varepsilon_m=0$ to get an elastic equation. It is much more
  complicated than $H=0$ and the catenoid solution indeed holds.}
\bibitem[{cat()}]{cat}
\bibinfo{note}{The first order correction to this solution for large but finite
  $r_s$ also involves a nonlinear differential equation and can not be solved
  analytically.}
\bibitem[{nec()}]{neck}
\bibinfo{note}{Intuitively, one may think that at $\alpha$ close to $\pi$, the
  {``neck"} of the membrane (see the lower-right panel of Fig.~\ref{fig:cat}
  for an illustration of the neck) cost a large elastic energy. In fact, this
  is not the case. In the soft membrane regime, the bending energy dominates.
  This neck can take a catenoid shape which has zero curvature energy. In the
  stiff membrane regime, the stretching energy dominates. The membrane can make
  a sharp turn to minimize the stretching energy again to almost zero.}
\end{thebibliography}

\end{document}